\documentclass[letter,twocolumn]{jpsj3}
\usepackage{txfonts}

\usepackage{amssymb}
\usepackage{bm}
\usepackage{url}

\usepackage{amsmath}
\usepackage{graphicx}
\usepackage{color}

\makeatletter
\newcommand\newblock{\hskip .11em\@plus.33em\@minus.07em}
\makeatother
\title{Coherence Length of Electronic Nematicity in Iron-Based Superconductors}

\author{Yoichi Kageyama$^{1,2,3}$, Asato Onishi$^{1,2,3}$, C\'edric Bareille$^{2,\ast}$, Kousuke Ishida$^{1,\dag}$, Yuta Mizukami$^{1,\ddag}$, Shigeyuki Ishida$^{3}$, Hiroshi Eisaki$^{3}$, Kenichiro Hashimoto$^{1}$, Toshiyuki Taniuchi$^{1,2,4}$, Shik Shin$^{5}$, Hiroshi Kontani$^{6}$, and Takasada Shibauchi$^{1}$}
\inst{
$^{1}$Department of Advanced Materials Science, The University of Tokyo, Kashiwa, Chiba 277-8561, Japan
\\$^{2}$The Institute for Solid State Physics (ISSP), The University of Tokyo, Kashiwa, Chiba 277-8581, Japan,
\\$^{3}$Research Institute for Advanced Electronics and Photonics, National Institute of Advanced Industrial Science and Technology (AIST), Tsukuba, Ibaraki 305-8568, Japan
\\$^{4}$Material Innovation Research Center (MIRC), The University of Tokyo, Kashiwa, Chiba 277-8561, Japan,
\\$^{5}$Office of University Professor, The University of Tokyo, Kashiwa, Chiba 277-8568, Japan
\\$^{6}$Department of Physics, Nagoya University, Furo-cho, Nagoya 464-8602, Japan
\\$^{\ast}$Present address: Hitachi High-Tech Corporation, Hitachinaka, Ibaraki 312-8504, Japan
\\$^{\dag}$Present address: Institute for Materials Research, Tohoku University, Sendai 980-8577, Japan
\\$^{\ddag}$Present address: Department of Physics, Tohoku University, Aoba-ku, Sendai 980-8578, Japan
}
\date{\today}

\abst{
  Recent developments in laser-excited photoemission electron microscopy (laser-PEEM) advance the visualization of electronic nematicity and nematic domain structures in iron-based superconductors. In FeSe and BaFe$_2$(As$_{0.87}$P$_{0.13}$)$_2$ superconductors, it has been reported that the thickness of the electronic nematic domain walls is unexpectedly long, leading to the formation of mesoscopic nematicity wave [T. Shimojima {\it et al.}, Science {\bf 373} (2021) 1122]. This finding demonstrates that the nematic coherence length $\xi_{\rm nem}$ can be decoupled from the lattice domain wall. Here, we report that the electronic domain wall thickness shows a distinct variation in related materials: it is similarly long in FeSe$_{0.9}$S$_{0.1}$ whereas it is much shorter in undoped BaFe$_2$As$_2$. We find a correlation between the thick domain walls and the non-Fermi liquid properties of normal-state resistivity above the nematic transition temperature. This suggests that the nematic coherence length can be enhanced by underlying spin-orbital fluctuations responsible for the anomalous transport properties.
}

\begin{document}
\maketitle

\begin{figure}[t]
  \centering{\includegraphics[width=\linewidth]{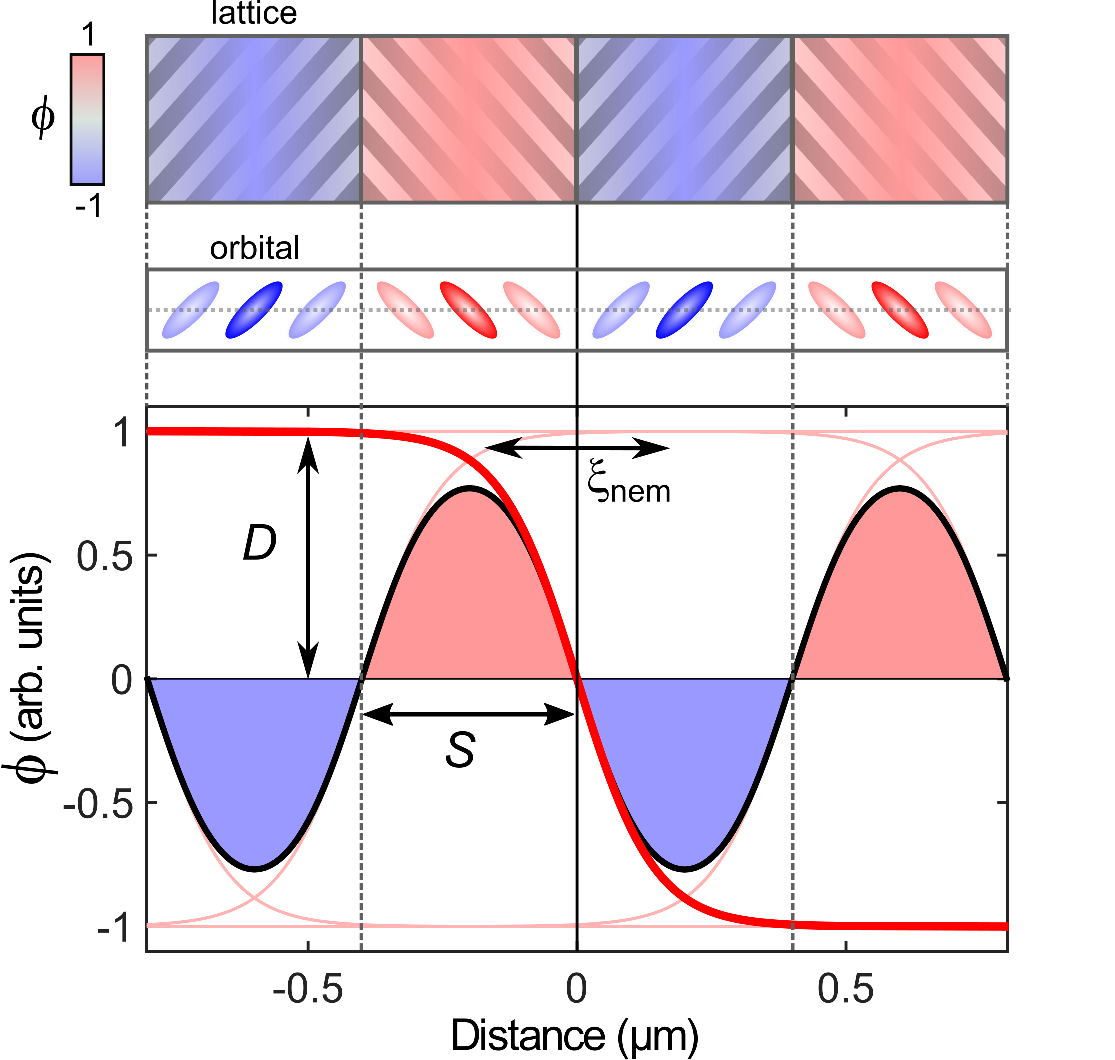}}
  \caption{Schematics of a mesoscopic nematicity wave. The upper panel sketches the lattice and orbital domains colored by the amplitude of the orbital polarization represented by the order parameter $\phi$. The stripes in the lattice and ellipsoids in the orbital domains represent the orthorhombic $a$ (or $b$) direction and $d_{xz}$ (or $d_{yz}$) dominated state, respectively. The lower panel simulates the mesoscopic nematicity wave (black line) of the order parameter $\phi$. The domain wall (red line) is characterized by the coherence length $\xi_{\mathrm{nem}}$, and the waveform is set by the wall interval $S$. Here we use $\xi_{\mathrm{nem}} = 450$ nm, $S = 400 $ nm, and the height of order parameter $D = 1$. }
  \label{fig:shematic_diagram}
\end{figure}

The electronic nematic phase which spontaneously breaks rotational symmetry has attracted great interest in strongly correlated materials such as cuprate \cite{Sato2017,auvrayNematicFluctuationsCuprate2019, Murayama2019,Ishida2020,nakataNematicityCuprateSuperconductor2021} and iron-based superconductors \cite{chuDivergentNematicSusceptibility2012,kasaharaElectronicNematicityStructural2012a}. In iron-based superconductors, the instability of the nematicity coupled with spin, orbital, and lattice degrees of freedom is considered to be closely related to unconventional superconductivity \cite{yoshizawaStructuralQuantumCriticality2012b, bohmerNematicSusceptibilityHoleDoped2014b, wangStrongInterplayStripe2016, kuoUbiquitousSignaturesNematic2016, ishidaPureNematicQuantum2022,Mukasa2023}. The nematic phase is usually accompanied by the long-range antiferromagnetic order in iron pnictides \cite{fernandesWhatDrivesNematic2014}, but a pure nematic phase without magnetic order can be found in iron chalcogenides \cite{Shibauchi2020}. 
Near the transition to the nematic phase, the divergent behaviors of nematic susceptibility are observed by several experimental probes \cite{Boehmer2022}, indicating the spin/orbital origins of the nematicity. Additionally, the coupling between the electrons and the lattice distorts the crystal structure from tetragonal to orthorhombic below the transition temperature $T_{\rm s}$. 
The nematic order can be characterized via two aspects: the anisotropy of the $t_{2g}$ orbitals (mainly $d_{xz}$ and $d_{yz}$ components) and the lattice distortion within the iron planes. 

\begin{figure*}[h]
  \center
      \includegraphics[width=\linewidth]{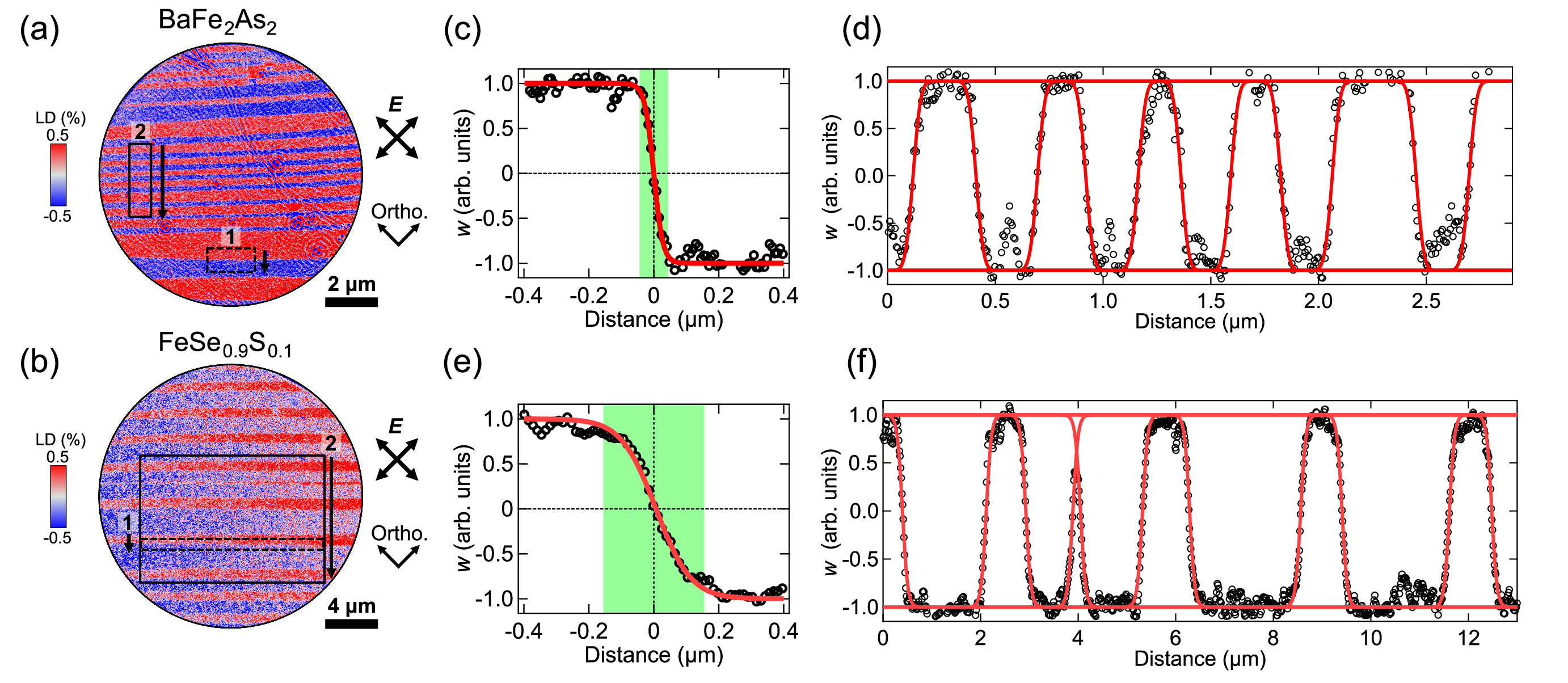}
      \caption{LD images in the electronic nematic states of iron-based superconductors. (a) LD image in BaFe$_2$As$_2$ at 55 K and (b) in FeSe$_{0.9}$S$_{0.1}$ at 45 K measured by linearly polarized light along the orthorhombic axes (crossed arrows). (c) Spatial dependence of normalized order parameter $w$ in BaFe$_2$As$_2$ obtained by averaging LD signals in area 1 of (a) near a domain boundary (dashed square). (d) Similar data obtained in area 2 of (a) containing several boundaries (solid square). (e) Spatial dependence of normalized order parameter $w$ in FeSe$_{0.9}$S$_{0.1}$ obtained by averaging LD signals in area 1 of (b) near a domain boundary (dashed square). (f) Similar data obtained in area 2 of (b) containing several boundaries (solid square). The data (open circles) in (c)-(f) obtained by averaging LD signals are plotted as a function of the distance along the arrows in (a) and (b). Red lines in (c)-(f) represent fitting results to Eq.\ (2). The domain walls are highlighted in green shades in (c) and (e). }
      \label{fig:domain_wall_thickness}
\end{figure*} 

These in-plane anisotropies lead to the formation of nematic domains, and the studies of the domain boundaries at which nematic direction switches are important to discuss the coherency of such nematicity. The polarized light microscopy and x-ray diffraction measurements reported the
observation of domain stripes along the tetragonal $[100]$ and $[010]$ directions,
corresponding to two different pairs of lattice distortion patterns \cite{tanatarDirectImagingStructural2009a}. The domain wall thickness of the orthorhombic lattice measured by the scanning tunneling spectroscopy is around 5 nm \cite{watashigeEvidenceTimeReversalSymmetry2015a}, consistent with the correlation length of lattice nematicity measured through soft phonon modes\cite{merrittNematicCorrelationLength2020a}. On the other hand, the real-space imaging of orbital anisotropy has become possible by laser-PEEM measurements recently, which reveal that the domain wall thicknesses of orbital nematic orders in FeSe and BaFe$_2$(As$_{0.87}$P$_{0.13}$)$_2$ are $\sim100$ times longer than those of the orthorhombic lattice \cite{shimojimaDiscoveryMesoscopicNematicity2021}. This exotic property can give rise to the peculiar real-space structure called mesoscopic nematicity wave \cite{shimojimaDiscoveryMesoscopicNematicity2021}, a sinusoidal wave of the nematic order parameter characterized by a period ranging from 400 to 1300 nm. As depicted in Fig.\,\ref{fig:shematic_diagram}, this nematicity wave forms when the domain size $S$ becomes comparable to the nematic coherence length $\xi_{\mathrm{nem}}$, which is the characteristic length scale of orbital switching. Here $\phi$ is the nematic order parameter, from which the height $D$ and the interval $S$ directly defines the period and amplitude of the stripe waves, with variation over different areas of the sample.

The presence of such a mesoscopic nematicity wave and the significantly longer coherence length of orbital nematic order compared to the lattice domain wall thickness suggest that electronic nematicity can decouple from lattice distortions near domain boundaries. This observation aligns with the recent results reporting an apparent absence of lattice distortion in the nematic phase of FeSe thin films \cite{Kubota2022}. To understand the mesoscopic nematicity wave, some theoretical proposals have been put forward. One scenario suggests that electron-electron interactions could greatly enhance the nematic coherence length \cite{onariOriginDiverseNematic2019b, shimojimaDiscoveryMesoscopicNematicity2021,Tazai2023}. Another possibility is the formation of an electronic density wave state, such as a surface smectic state, which might manifest as an apparent thickening of the domain wall \cite{lahiriDefectinducedElectronicSmectic2021b}. Despite these discussions, the precise origin of the mesoscopic nematicity wave remains unclear.

In this study, we use the linear dichroism (LD) by laser-PEEM \cite{t.taniuchiUltrahighspatialresolutionChemicalMagnetic2015} to visualize the nematic states of two iron-based superconductors, BaFe$_2$As$_2$ (Ba122), and FeSe$_{0.9}$S$_{0.1}$, which are sister materials of the ones (BaFe$_2$(As$_{0.13}$P$_{0.87}$)$_2$ and FeSe) studied in the previously reported laser-PEEM measurements \cite{shimojimaDiscoveryMesoscopicNematicity2021}. We employ LD asymmetry $(I^+ - I^-)/(I^+ + I^-) $ to evaluate the LD signal where $I^{+}$ and $I^{-}$ are the photoemission intensities of linear polarized light perpendicular to each other. The light source of laser-PEEM is a continuous-wave laser of 4.66 eV with an incidence normal to the sample surface. By using this low-energy light source, we only measure electronic state near the Fermi-level and at the vicinity of the Brillouin zone center, which is here dominated by the $t_{2g}$ orbitals \cite{t.taniuchiUltrahighspatialresolutionChemicalMagnetic2015,Taniuchi2016}. The planar shape of these orbitals leads to a strong suppression of the electron excitation when the light polarization is normal to the orbital plane \cite{zhangSymmetryBreakingOrbitaldependent2012}. We can then visualize the amplitude of the orbital polarization in real space by mapping the LD asymmetry. Here, we report that orbital nematic domains of undoped Ba122 are more like a square wave rather than a sinusoidal wave as reported in the previous study of BaFe$_2$(As$_{0.87}$P$_{0.13}$)$_2$, and the domain wall thicknesses are much shorter in the undoped case. In FeSe$_{0.9}$S$_{0.1}$, however, we find a similarly thick domain wall as the previous results in FeSe. Comparisons with the transport data above $T_{\rm s}$ reveal that this difference in nematic domain wall thicknesses between Ba122 and FeSe correlates with the power-law exponent of the temperature dependence of resistivity, which is related to the deviations from the Fermi-liquid properties. These results point to the importance of spin-orbital fluctuations responsible for the non-Fermi liquid transport on the coherence length of the electronic nematic order.

Figures \ref{fig:domain_wall_thickness}(a) and (b) show the LD images obtained for the cleaved surfaces of Ba122 and FeSe$_{0.9}$S$_{0.1}$, respectively. In both cases, clear stripe patterns of LD asymmetry are resolved, which signifies the domain structures of the nematic orbital order below $T_{\rm s}$. The LD signal was taken by the difference in the photoemission intensities of parallel and perpendicular polarized light to the orthorhombic axis. To evaluate the domain wall thickness, we analyze the data in areas where the domain wall interval $S$ is wide enough. We use the Ginzburg-Landau (GL) theory, in which the free energy of the system can be written as \cite{shimojimaDiscoveryMesoscopicNematicity2021}
\begin{flalign}
F=\int [ a\phi(\bm{r})^2+\frac{b}{2}\phi(\bm{r})^4+c|\nabla \phi(\bm{r})|^2 ] d\bm{r},
\label{eq:GL_freeenergy}
\end{flalign}
where $\phi (\bm{r})$ is the nematic order parameter, the coefficient $a$ is proportional to $T-T_{\rm s}$, and both $b$ and $c$ are positive coefficients. Near a domain boundary along $y$ direction at $x=0$, the $x$-dependence of the order parameter is given by
\begin{flalign}
\phi(x)=D \tanh(x/2r_c),
\label{eq:domain_wall_func}
\end{flalign}
where $D=\sqrt{{|a|}/{b}}$, and $r_c=\sqrt{{2c}/{|a|}}$. Here we define the nematic coherence length $\xi_{\mathrm{nem}}$ as a characteristic domain wall thickness $\xi_{\mathrm{nem}}=2\pi r_c$, which corresponds to the distance between the points where $\phi(x)$ changes from $\sim -0.92D$ to $\sim 0.92D$ as shown in Fig.\ \ref{fig:shematic_diagram} (or equivalently, from $0$ to $\sim 0.996D$) \cite{shimojimaDiscoveryMesoscopicNematicity2021}. We note that the nematic coherence length is related to the coefficient $c$ in the third term of Eq.\ (\ref{eq:GL_freeenergy}), which determines the stiffness of the electronic nematicity. In other words, when the nematic order parameter is hard to change as a function of position, the coherence length becomes long.

To analyze the LD images, we have to take into account the spatial resolution of the PEEM image, which is estimated as $\lesssim 92$ nm from the analysis using crystal defects \cite{shimojimaDiscoveryMesoscopicNematicity2021}. 
The actual observed LD signal $\Phi(x)$ is convoluted by the Gaussian with the standard deviation $\sigma$ as 
\begin{flalign}
\Phi (x)= \frac{1}{\sigma\sqrt{\pi}} \int \phi(x-t)\exp\left(-\frac{t^2}{\sigma^2}\right) dt, 
\label{eq:convolution}
\end{flalign} 
where $\phi(x)$ is the order parameter shown in Fig.\ \ref{fig:shematic_diagram} and Eq.\ (\ref{eq:domain_wall_func}), and $2\sigma\sqrt{2\log 2}$ corresponds to the spatial resolution, which is the full width at half maximum of the Gaussian. 
In Fig.\ \ref{fig:domain_wall_thickness}(c), we plot the spatial dependence of the normalized LD order parameter $w = \Phi(x)/D$ near a domain boundary in Ba122. In this material, the LD signal changes sign sharply near the boundary, and the spatial dependence of $w$ can be fitted nicely with Eq.\ (\ref{eq:convolution}). However, we find that the nematic coherence length is resolution limited, namely $\xi_{\mathrm{nem}}\lesssim  92$ nm. We confirm this resolution-limited changes for several domain walls as shown in Fig.\ \ref{fig:domain_wall_thickness}(d). In the previous optical measurements of Ba122 systems, the reported typical domain size is around a few hundred nm to a few $\mu$m \cite{tanatarDirectImagingStructural2009a, maMicrostructureTetragonaltoorthorhombicPhase2009a, thewaltImagingAnomalousNematic2018a}, which results in the condition $S\gg \xi_{\mathrm{nem}}$. Correspondingly, we only observe square-wave-like domain structures rather than a sinusoidal wave. 

In contrast, the LD signal in FeSe$_{0.9}$S$_{0.1}$ shows a much broader change at the boundary as shown in Figs.\ \ref{fig:domain_wall_thickness}(e) and (f). Here we fit the data for 11 domain walls simultaneously by varying the interval $S$. As a result, all the data can be fitted successfully by taking the coherence length $\xi_{\mathrm{nem}} =310$ nm as the fitting parameter. This value is not resolution limited, which indicates that the orbital nematic coherence length is much longer than the typical lattice domain wall thickness, as previously found in FeSe and BaFe$_2$(As$_{0.87}$P$_{0.13}$)$_2$ \cite{shimojimaDiscoveryMesoscopicNematicity2021}.
Combining these results with the previous work, we summarize the nematic coherence length $\xi_{\mathrm{nem}}$ for four iron-based superconductors in Table \ref{table:summary_of_thickness}. 
These results show that the coherence length of electronic nematic order varies in these iron-based superconductors. We note that the nematic phase accompanies antiferromagnetism in Ba122 systems but not in FeSe-based materials, implying that the order of spin state is not the key parameter for the coherence length. We also note that the domain wall thickness does not change significantly as a function of temperature except near $T_{\rm s}$, where the analysis becomes difficult \cite{shimojimaDiscoveryMesoscopicNematicity2021}. This apparently contradicts the temperature dependent $\xi_{\mathrm{nem}}$ expected in the simple GL free energy discussion based on Eq.\ (\ref{eq:GL_freeenergy}). However, this may be related to the weakly first-order transition at $T_{\rm s}$ found in some iron-based superconductors \cite{fernandesWhatDrivesNematic2014}, which requires additional higher-oder terms of the order parameter in the free energy. In such a case, the divergence of $\xi_{\mathrm{nem}}$ can be avoided. Moreover, the apparent temperature-independent behavior of coherence length can be explained by the strong-coupling effect of the nematic order, which flattens the temperature dependence of energy gap ($\propto 1/\xi_{\mathrm{nem}}$) except near the transition temperature.
%This suggests that additional pinning mechanisms may exist in determining the domain structures.     

\begin{table}[b]
  \caption{Domain wall height $D$ and nematic coherence length $\xi_{\rm nem}$ obtained from the LD images and the nematic transition temperatures $T_{\rm s}$ for four iron-based superconductors.}
  \label{table:summary_of_thickness}
  \begin{tabular}{lcccc}
    \hline
     \ \ \ \ \ \  &\ \ $D$ (arb. units) \ \ &\ \ $\xi_{\mathrm{nem}}$ (nm)\ \ &\ \ $T_{\mathrm{s}}$ (K) \\ %& \\ $T_c$ \ \   \\
    \hline \hline
    BaFe$_2$As$_2$ &0.50 &$\lesssim  92$  & 138 \\
    BaFe$_2$(As$_{0.87}$P$_{0.13}$)$_2$ \cite{shimojimaDiscoveryMesoscopicNematicity2021} &0.13 & 450 & 93 \\
    FeSe  \cite{shimojimaDiscoveryMesoscopicNematicity2021} & 0.11& 550 & 90 \\
    FeSe$_{0.9}$S$_{0.1}$ & 0.29 & 310  & 78 \\
    \hline
  \end{tabular}
\end{table}

\begin{figure}[t]
    \centering\includegraphics[width=0.9\linewidth]{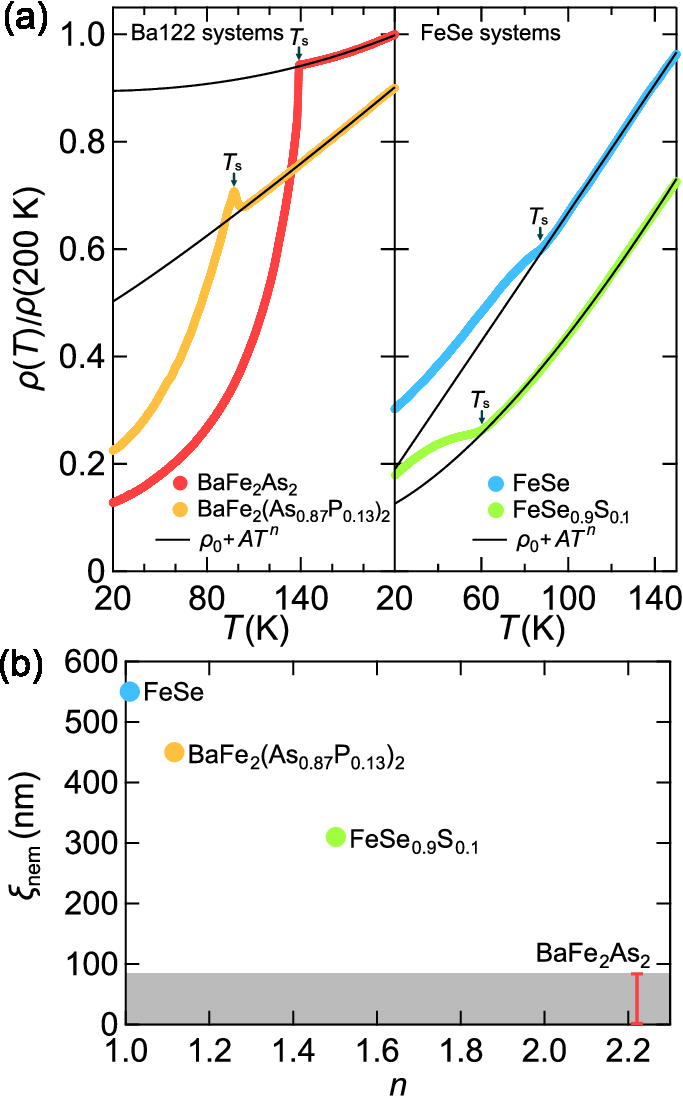}
    \caption{Comparisons of transport properties in four iron-based superconductors. (a) Temperature dependence of in-plane resistivity $\rho(T)$ at zero field normalized at the value at 200 K. The data for BaFe$_2$(As$_{0.13}$P$_{0.87}$)$_2$ is from Ref.\ \cite{Nakajima2012} and shifted vertically by $-0.1$ and the data for FeSe is from Ref.\ \cite{Sun2015} and shifted vertically by $+0.2$ for clarity. The normal-state resistivity curves above $T_{\rm s}$ (arrows) are fitted to the power-law dependence $\rho_0+AT^n$ (black lines). (b) Nematic coherence length $\xi_{\mathrm{nem}}$ obtained in the LD analysis as a function of the power-law exponent $n$. The gray shaded area is below the spatial resolution of our measurements.}
    \label{fig:resistivity}
\end{figure} 

To gain some insights into the observed unusual material variation of nematic coherence length, we compare the temperature dependence of resistivity $\rho(T)$ in these iron-based superconductors, as shown in Fig.\ \ref{fig:resistivity}(a). The $\rho(T)$ curves show noticeable anomalies at $T_{\rm s}$, below which the resistivity goes up or down depending on the competition between changes in the density of states and scattering rate. We fit the data above the nematic transition $T_{\rm s}$ to the simple power-law dependence $\rho(T) =\rho_0 + AT^n$, where $\rho_0$ is the residual resistivity, $A$ is the coefficient, and $n$ is the temperature exponent. We find that the exponent $n$ varies strongly in these materials, ranging from $n\sim 2$ expected for the standard Fermi-liquid theory of metals to $n\sim 1$, a hallmark of non-Fermi liquid properties. Comparisons between this temperature exponent $n$ and nematic coherence length $\xi_{\mathrm{nem}}$ are indicated in Fig.\ \ref{fig:resistivity}(b), showing a striking correlation between the non-Fermi liquid properties and long $\xi_{\mathrm{nem}}$ values. The results suggest that the source of deviations from the Fermi-liquid properties above $T_{\rm s}$, which is most likely spin-orbital fluctuations that are relevant to the nematic ordering at $T_{\rm s}$, plays a significant role in enhancing the coherence length of electronic nematic order.

Recent electron-correlation theories based on Aslamazov-Larkin vertex corrections \cite{onariOriginDiverseNematic2019b,Tazai2023,Onari2012} suggest that the electronic nematic order can be driven by the multimode interference effects of spin-orbital fluctuation channels. This highlights the importance of antiferroic fluctuations in stabilizing the nematic orbital order, in which the dynamic susceptibility $\chi(\bm{q})$ is governed by the term $(\bm{q}-\bm{Q})^{-1}$ where $\bm{Q}$ is the nesting vector \cite{onariOriginDiverseNematic2019b}. It would be then expected that when such fluctuations that drive the nematic order are strong enough to give rise to deviations from the Fermi-liquid behavior, the nematic order may have high stiffness, which results in a long coherence length. 
Within this framework, the GL coefficients for the nematic order, $a\equiv {\dot a}(T-T_{\rm s})$ and $c$ in Eq. (\ref{eq:GL_freeenergy}), have been studied theoretically \cite{Tazai2023}. Here ${\dot a} \ (>0)$ is the temperature slope of $a(T)$ at the transition temperature. The GL theory for Bardeen-Cooper-Schrieffer (BCS) superconductors gives the relations ${\dot a}\sim N(0)T_{\rm c}^{-1}$ and $c\sim N(0)(\bm{K}/2)^{-2}$, where $N(0)$ is the density of states at the Fermi level, $T_{\rm c}$ is the critical temperature, and $\bm{K}$ is the reciprocal vector.
For the nematic order, in contrast, the relations ${\dot a}\ll N(0)T_{\rm s}^{-1}$ and $c\gg N(0)(\bm{K}/2)^{-2}$ can be derived \cite{Tazai2023}. The obtained small ${\dot a}$ explains %the wide nematic critical region and 
the small specific heat jump at $T_{\rm s}$ in iron-based superconductors. The large $c$ coefficient originates from the good nesting between the hole and electron Fermi-surface pockets. Therefore, the nematic domain wall thickness $\xi_{\rm nem}\propto \sqrt{2c/|a|}$ can become much longer than the BCS coherence length $\xi_{\rm BCS}\sim v_{\rm F}/T_{\rm c}$ with the Fermi velocity $v_{\rm F}$. Thus, the nematic coherence length can be strongly enhanced by the underlying antiferroic fluctuations, which are related to the Fermi-surface nesting.  
%The observed systematic change in the nematic domain wall thickness provides useful information to elucidate the essential electronic properties of Fe-based superconductors.

Iron-based superconductors exhibit multiband electronic structures with separated hole and electron Fermi surfaces, which are connected essentially by $\bm{Q}\approx (\pi, \pi)$ in the two-Fe unit-cell notation. However, the details of the nesting properties and spin-orbital fluctuations depend on several factors, such as the sizes of hole and electron pockets, the three-dimensional warping of Fermi surfaces, and the momentum dependence of orbital components in each Fermi surface. Therefore, it is nontrivial to predict the nematic coherence length for each material, and the full understanding of the observed variation in $\xi_{\mathrm{nem}}$ requires further theoretical studies. Nevertheless, the present finding of the correlation between $\xi_{\mathrm{nem}}$ and the non-Fermi liquid transport supports the above scenario of the importance of underlying spin-orbital fluctuations on the nematicity stiffness.

In summary, we have presented our laser-PEEM imaging studies of the electronic nematic domains in iron-based superconductors, from which the nematic coherence length $\xi_{\mathrm{nem}}$ is evaluated. We find that the domain wall thickness in the electronic nematic states shows a material variation in iron-based superconductors, which correlates with the deviations from the Fermi-liquid properties in the electronic transport above the nematic transition temperature. This correlation highlights the relationship between the stiffness of nematic order and spin-orbital fluctuations. Our results also shed light on the importance of real-space observation with a mesoscopic scale to understand the quantum liquid crystal states of strongly correlated materials.

We thank fruitful discussions with Takahiro Shimojima. This work was supported by Grants-in-Aid for Scientific Research (KAKENHI) (No.\,JP22H00105), Grant-in-Aid for Scientific Research on innovative areas ``Quantum Liquid Crystals'' (No.\,JP19H05824) and Grant-in-Aid for Scientific Research for Transformative Research Areas (A) ``Condensed Conjugation'' (No.\,JP20H05869) from Japan Society for the Promotion of Science (JSPS).

\bibliographystyle{jpsj} 
\bibliography{PEEM_ref}

\end{document}